\def\slash#1{\setbox0=\hbox{$#1$}#1\hskip-\wd0\dimen0=5pt\advance
       \dimen0 by-\ht0\advance\dimen0 by\dp0\lower0.5\dimen0\hbox
         to\wd0{\hss\sl/\/\hss}}
\documentstyle[12pt]{article}

\newlength{\dinwidth}
\newlength{\dinmargin}
\newcommand{\resection}[1]{\setcounter{equation}{0}\section{#1}}

\begin{document}

\def\lq{\left [}
\def\rq{\right ]}
\def\LL{{\cal L}}
\def\VV{{\cal V}}
\def\AA{{\cal A}}
\def\gi{{g_{P^* P \pi}}}
\def\gis{{g_{P^* P^* \pi}}}
\def\lq{\left [}
\def\rq{\right ]}
\def\qq{<{\overline u}u>}
\def\dmu{\partial_{\mu}}
\def\dmus{\partial^{\mu}}

\def\gid{{g_{D^* D \pi}}}
\def\gib{{g_{B^* B \pi}}}
\def\gids{{g_{D^* D^* \pi}}}
\def\gibs{{g_{B^* B^* \pi}}}
\def\mbs{{m_{B^*}}}
\def\fbs{{f_{B^*}}}

\newcommand{\be}{\begin{equation}}
\newcommand{\ee}{\end{equation}}
\newcommand{\bea}{\begin{eqnarray}}
\newcommand{\eea}{\end{eqnarray}}
\newcommand{\nn}{\nonumber}
\newcommand{\dd}{\displaystyle}
\vspace*{1cm}
\begin{center}
  \begin{Large}
  \begin{bf}
HEAVY MESON HYPERFINE SPLITTING: A COMPLETE $1/m_Q$ CALCULATION.
\footnote{Partially supported by the Swiss National Foundation.}\\
  \end{bf}
  \end{Large}
  \vspace{5mm}
  \begin{large}
N. Di Bartolomeo and R. Gatto\\
  \end{large}
D\'epartement de Physique Th\'eorique, Univ. de Gen\`eve\\
  \vspace{5mm}
  \begin{large}
F. Feruglio\\
  \end{large}
Dipartimento di Fisica, Univ.
di Padova\\
I.N.F.N., Sezione di Padova\\
  \vspace{5mm}
  \begin{large}
G. Nardulli\\
  \end{large}
Dipartimento di Fisica, Univ.
di Bari\\
I.N.F.N., Sezione di Bari\\
  \vspace{5mm}
\end{center}
  \vspace{2cm}
\begin{center}
UGVA-DPT 1994/10-864\\\
BARI/TH 190-94\\\
hep-ph/9411210 \\\
October 1994
\end{center}
\vspace{1cm}
\noindent
$^*$ Partially supported by the Swiss National Foundation
\newpage

\begin{quotation}
\vspace*{5cm}
\begin{center}
  \begin{large}
  \begin{bf}
  ABSTRACT
  \end{bf}
  \end{large}
\end{center}
  \vspace{5mm}
\noindent

The hyperfine splittings
$\Delta_D=(m_{D^*_s}-m_{D_s})-(m_{D^{*+}}-m_{D^+})$
and
$\Delta_B=(m_{B^*_s}-m_{B_s})-(m_{B^{*0}}-m_{B^0})$
are analyzed in the framework
of an effective lagrangian
possessing chiral, heavy flavour and spin symmetries,
explicitly broken by a complete set of first order terms.
Among these terms, those responsible for
the difference between the couplings $\gis$ and $\gi$ are
evaluated in the QCD sum rules approach.
Their contribution to $\Delta_D$ and to $\Delta_B$ appears to
quantitatively balance previously estimated chiral effects
nice agreement with the experimental data, solving a suspected puzzle for heavy
quark theory.

\end{quotation}

\newpage
\setcounter{page}{1}
\resection{Introduction}
The spectroscopy of heavy mesons is among the simplest framework where the
ideas and the
methods of heavy quark expansion can be quantitatively tested.
Recently, attention has been focused on the combinations
\cite{ros,ran,jen,chi}:
\be
\Delta_D=(m_{D^*_s}-m_{D_s})-(m_{D^{*+}}-m_{D^+})
\label{a}
\ee
\be
\Delta_B=(m_{B^*_s}-m_{B_s})-(m_{B^{*0}}-m_{B^0})
\label{b}
\ee
which are measured to be \cite{PDB}:
\be
\Delta_D\simeq 1.0\pm1.8~MeV
\label{c}
\ee
\be
\Delta_B\simeq 1.0\pm2.7~MeV
\label{d}
\ee
The above hyperfine splitting is free from electromagnetic corrections and it
vanishes
separately in the $SU(3)$ chiral limit and in the heavy quark limit. In the
combined
chiral and heavy quark expansion, the leading contribution is of order
$m_s/m_Q$ and
one would expect the relation \cite{ros}:
\be
\Delta_B=\frac{m_c}{m_b}\Delta_D
\label{e}
\ee
In the so called heavy meson effective theory \cite{wis}, which combines
the heavy quark expansion and the chiral symmetry, there is only one
lowest order operator
contributing to $\Delta_{D,B}$:
\be
\lambda_2 {\cal O}_2=\frac{\lambda_2}{8}
Tr[{\bar H}^i_a \sigma_{\mu\nu} H^b_j \sigma^{\mu\nu}]
                     (m_Q^{-1})^j_i\frac{(m_{\xi})^a_b}{\Lambda_{CSB}}
\label{f}
\ee
where $i$, $j$ are heavy flavour indices and  $a$,$b$ light flavour indices.
The $4 \times 4$ Dirac matrix $H^i_a$ describes the spin doublet $P$, $P^*$,
with $P$ heavy meson composed by the heavy quark $Q_i$ and the light antiquark
${\bar q}_a$. The matrix $m_{\xi}$ is
\be
m_{\xi}=(\xi m_q \xi +\xi^\dagger m_q \xi^\dagger)
\label{g}
\ee
Here $m_q$ is the light quarks mass matrix and $\xi =exp(iM/f)$, where
$M$ is the pseudoscalar $3 \times 3$ matrix and $f$ the pseudoscalar decay
constant (we take $f=132 \; MeV$).

By taking $m_s/\Lambda_{CSB}\simeq 0.15$ and by taking $\lambda_2 \simeq
\Lambda_{QCD}^2 \simeq 0.1 GeV^2$
one would estimate:
\be
\Delta^{(2)}_D\simeq 20~MeV
\label{h}
\ee
\be
\Delta^{(2)}_B\simeq 6~MeV
\label{i}
\ee
Given the present experimental accuracy,
the above estimate is barely acceptable, as an order of magnitude,
for $\Delta_B$,
while it clearly fails to reproduce the data for $\Delta_D$.
If the contribution from ${\cal O}_2$ were the only one
 responsible for the hyperfine
splittings, agreement with the data clearly would require
 a rather small value for $\lambda_2$.

In chiral perturbation theory, an independent contribution arises from
one-loop corrections to the heavy meson self energies \cite{jen},
evaluated from an initial lagrangian containing, at the lowest order,
both chiral breaking and spin breaking terms. The loop corrections
in turn depend on an arbitrary renormalization point $\mu^2$ (e.g. the
t'Hooft mass of dimensional regularization). This dependence is cancelled by
the $\mu^2$ dependence of the counterterm $\lambda_2(\mu^2) {\cal O}_2$,
as it should happen for any physical result. A commonly accepted
point of view is that the overall effect of adding the counterterm
consists in replacing $\mu^2$ in the loop corrections with the physical scale
relevant to the
problem at hand, $\Lambda_{CSB}^2$. Possible finite terms in the counterterm
are supposed to be small compared to the large chiral logarithms.
With this philosophy in mind,
two classes of such corrections has been estimated in ref. \cite{ran}
and they give (for the values of the parameters given by these authors):
\be
\Delta^0_D\simeq +30~MeV,~~~~~\Delta^1_D\simeq +65~MeV,
\label{l}
\ee
\be
\Delta^0_B\simeq + 10~MeV,~~~~~\Delta^1_B\simeq +22~MeV,
\label{m}
\ee
Here $\Delta^0$ represents the contribution of the chiral logarithm
and $\Delta^1$ is a non analytic contribution, of order $m_s^{3/2}$.
This provides a rather uncomfortable situation since, to account for the
observed data, one should require an accurate and
innatural cancellation between $(\Delta^0+\Delta^1)$
and the finite terms from $\Delta^{(2)}$, contrary to the usual expectation.

The chiral computation giving $\Delta^0+\Delta^1$ is however incomplete
\cite{chi}, because it does not include the spin breaking
effect due to the difference between the $P^* P^* \pi$
and the $P^* P \pi$ couplings ($P=D,B$), defined by the relations:
\be
<\pi^-(q)~P^o(q_2) | P^{*-}(q_1,\epsilon )> \; = 2 \; \gi \; \frac{m_P}{f_\pi}
\epsilon^{\mu} \cdot q_{\mu}
\label{sda}
\ee
\be
<\pi(q)~P^{*}(q_2, \epsilon_2) | P^{*}(q_1,\epsilon_1 )> \; = \; -i
\frac{2}{f_{\pi}} \gis \; \epsilon_{\mu\nu\alpha\beta} \epsilon_1^{\mu}
\epsilon_2^{\nu} q^{\alpha} q_1^{\beta}
\label{1}
\ee
The scaling law $2 g_{P^* P \pi }m_P/f_{\pi}$ for the strong
$D^* D \pi$ coupling constant was first proposed in \cite{pham,nussi}.
The splitting between the coplings (\ref{sda}) and (\ref{1}) is
of order $1/m_Q$, and therefore has to be taken into account in the
chiral computation, to work consistently at the desired order.

In the present paper we will provide an estimate of $g_{P^* P^* \pi}-g_{P^* P
\pi}$
based on a QCD sum rule, and, by including this additional spin breaking
effect,
we will complete the evaluation of $\Delta_{D,B}$ coming from the chiral loops.

\resection{The Hyperfine Splitting}

To better clarify the importance of $g_{P^*P^* \pi}-g_{P^* P \pi}$ for the
problem at hand,
we remind that the effective lagrangian for heavy mesons and light
pseudoscalars,
at first order in $m_Q^{-1}$ and in the light quark masses $m_q$ reads:
\be
{\cal L}={\cal L}_0+{\cal L}^q+{\cal L}^Q
\label{n}
\ee
Here ${\cal L}_0$ represents the chiral, heavy flavour and spin symmetric term:
\bea
{\cal L}&=&-iTr[{\bar H}^i_a v_\mu \partial^\mu H^a_i]+
           \frac{f^2}{8}Tr [\partial_\mu \Sigma^\dagger \partial^\mu
\Sigma]\nn\\
          &+&\frac{i}{2}Tr[{\bar H}^i_a  H^b_i] v^\mu
             (\xi^\dagger \partial_\mu\xi+\xi \partial_\mu\xi^\dagger)_b^a\nn
\\
          &+&\frac{i}{2}g Tr[{\bar H}^i_a  H^b_i \gamma_\mu\gamma_5] ({\cal
A}^\mu)_b^a
\label{o}
\eea
where $\Sigma=\xi^2$ and:
\be
{\cal A}_\mu=\xi^\dagger \partial_\mu\xi-\xi \partial_\mu\xi^\dagger
\label{p}
\ee

{}From the last term in eq. (\ref{o}), one obtains the $P^* P \pi$ and
$P^* P^* \pi$ couplings defined in eqs. (\ref{sda}) and (\ref{1}), in
the limit $m_P\to \infty$:
\be
\gi=\gis=g
\label{equal}
\ee

The leading chiral breaking corrections are given by:
\bea
{\cal L}^q&=&\lambda_0 Tr[m_q \Sigma + \Sigma^\dagger m_q]\nn \\
          &+&\lambda_1 Tr[{\bar H}^i_a  H^b_i](m_\xi)_b^a\nn \\
          &+&\lambda'_1 Tr[{\bar H}^i_a  H^a_i](m_\xi)_a^a
\label{q}
\eea
The second term in eq. (\ref{q}) is responsible
for the mass splitting between strange and
non-strange heavy mesons:
\be
\Delta_s = 2 \lambda_1 m_s
\ee
One has approximately
$\Delta_s \simeq 100 \; MeV$, $\lambda_1 \simeq 0.33$.

The third term, listed for completeness, gives an equal contribution
to each heavy meson mass,
it does not affect the hyperfine splitting, and it does not play any role in
our analysis.

Finally the terms of order $1/m_Q$, breaking either the heavy flavour or the
spin symmetries, are given by:

\bea
{\cal L}^Q&=&-\frac{\lambda}{8} Tr[{\bar H}^i_a \sigma_{\mu\nu} H^a_j
\sigma^{\mu\nu}]
             (m_Q^{-1})^j_i\nn\\
          &+&\frac{i g}{2}\frac{(a+b)}{2}
 Tr[{\bar H}^i_a  H^b_j \gamma_\mu\gamma_5]
              (m_Q^{-1})^j_i({\cal A}^\mu)_b^a \nn\\
          &+&\frac{i g}{2}\frac{(a-b)}{2}
 Tr[{\bar H}^i_a \gamma_{\mu}\gamma_5 H^b_j]
             (m_Q^{-1})^j_i ({\cal A}^{\mu})_b^a
\label{s}
\eea
The first term in eq. (\ref{s})
 is responsible for the splitting $\Delta$ between the
$1^-$ and $0^-$ heavy meson masses:
\be
\Delta=\frac{2\lambda}{m_Q}
\label{t}
\ee
For the $B,B^*$ system $\Delta\simeq 46~MeV$, whereas
for $D,D^*$ $\Delta$ is approximately $141~MeV$, so that
one has:
 \be \lambda\simeq 0.10-0.11~GeV^{2} \label{tbis} \ee
The second term in (\ref{s})
 breaks only the heavy flavour symmetry, making the
$B^* B^{(*)} \pi$ and $D^* D^{(*)}\pi$ couplings different.
The third term breaks also the spin symmetry and contributes
differently to the $P^* P \pi$ and to the $P^* P^* \pi$ couplings.
This is precisely the effect relevant to the hyperfine splitting.
To this order
in $1/m_Q$ one has:
\be
\gis= g \left( 1+\frac{a}{m_Q} \right) \; \;\;\;~~
\gi= g \left( 1+\frac{b}{m_Q} \right)  \label{17b}
\ee
and
\be
\Delta_g\equiv g_{P^*P^* \pi}-g_{P^* P \pi}=g \frac{a-b}{m_Q}
\label{u}
\ee
The chiral and spin symmetry breaking parameters relevant to the hyperfine
splitting
are the light pseudoscalar masses $m_\pi$, $m_K$ and $m_\eta$, $\Delta_s$,
$\Delta$ and
$\Delta_g$. In terms of these quantities, one finds \cite{jen,ran,chi}:
\bea
\Delta_{P}&=&\frac{g^2 \Delta}{16 \pi^2 f^2}
            \Bigl[4 m_K^2 ln(\frac{\Lambda_{CSB}^2}{m_k^2})+
                  2 m_\eta^2 ln(\frac{\Lambda_{CSB}^2}{m_\eta^2})-
                  6 m_\pi^2 ln(\frac{\Lambda_{CSB}^2}{m_\pi^2})\Bigr]\nn\\
          &+&\frac{g^2 \Delta}{16 \pi^2 f^2} [24 \pi m_K \Delta_s]\nn\\
          &-&\frac{g^2}{6 \pi f^2}\frac{\Delta_g}{g}
             (m_K^3+\frac{1}{2} m_\eta^3-\frac{3}{2} m_\pi^3)
\label{split}
\eea
The dependence upon the heavy flavour $P=D,B$ is contained in the parameters
$\Delta$
and $\Delta_g$.

The first term in eq. (\ref{split})
is the so called chiral logarithm
\cite{jen}.
In the ideal situation with pseudoscalar masses much smaller than
$\Lambda_{CSB}$,
it would represent the dominant contribution to $\Delta_P$. For the case of $D$
and
$B$ mesons the corresponding values have been listed in eqs. (\ref{l}) and
(\ref{m})
as $\Delta^0_D$ and
$\Delta^0_B$, respectively. There a value $g^2 =0.5$ has been used.

The second term in eq. (\ref{split}) represents a non analytic
contribution of order $m_s^{3/2}$ \cite{ran}, which, although formally
suppressed with respect to the leading one, is numerically more important,
because
of the large coefficient $24 \pi$. It is given
by $\Delta^1_D$ and $\Delta^1_B$ in eqs. (\ref{l}) and (\ref{m}).

Finally, the last term in eq. (\ref{split})
\cite{chi} is
also of order $m_s^{3/2}$. It can be numerically important as soon as
$\Delta_g/g$ is of order $10\%$ and, if equipped with the right sign,
it can cause
a substantial cancellations of the previous two contributions.

\resection{ QCD Sum Rules for $\gi$ and $\gis$}

The coupling $\gi$ has already been calculated in \cite{noi} by means of QCD
sum rules and here we proceed to a similar computation concerning the coupling
$\gis$.
We start from the correlator:
\be
A_{\mu\nu}(q_1,q) = i \int dx <\pi(q)| T(V_{\mu}(x) V_{\nu}^{\dagger}(0)
 |0> e^{-iq_1x} = A(q_1^2, q_2^2,q^2)\epsilon_{\mu\nu\alpha\beta} q^{\alpha}
q_1^{\beta}+ \ldots
\label{2}
\ee
where $V_{\mu}={\overline u} \gamma_{\mu} Q$  is the interpolating vector
current for the $P^*$ meson.

We compute the scalar function $A$ in the soft pion limit $q \to 0$. This
implies $q_1=q_2$ forcing to use a single Borel transformation,
and it is the origin of the so called parasitic terms \cite{noi}.
The correlator in (\ref{2})
can be calculated by an Operator Product Expansion: we keep all the operators
with dimension up to five, arising from the expansion of the current
$V_{\mu}(x)$ at the third order in power of $x$, and the heavy quark propagator
to the second order. The result is:
\bea
A(q_1^2,q_1^2,0) &= & \frac{f_{\pi}}{q_1^2-m_b^2}+
\frac{1}{(q_1^2-m_b^2)^2}\left[ {{2\qq m_b}\over{3 f_{\pi}}}+{{8 f_{\pi}
m_1^2}\over{9}} \right] \nn\\
&+& \frac{1}{(q_1^2-m_b^2)^3}\left[
 -{{10 m_b^2 f_{\pi} m_1^2}\over{9}}+{{m_0^2\qq m_b}
\over{3 f_{\pi}}}\right]+ \nn \\
& -&\frac{1}{(q_1^2-m_b^2)^4}
{{m_0^2\qq m_b^3}\over{ f_{\pi}}}
\label{3}
\eea
In eqs.(\ref{3}) $\qq$ is the quark condensate ($\qq =-(240 MeV)^3$), $m_0$
and $m_1$ are defined by the equations
\be
<{\overline u} g_s \sigma \cdot G u> =m_0^2\qq
\ee
\be
<\pi (q)|{\overline u} D^2 \gamma_{\mu} \gamma_5 d |0> = -i f_{\pi} m_1^2
q_{\mu}
\ee
and their numerical values are: $m_0^2=0.8 \; GeV^2$, $m_1^2=0.2 \; GeV^2$
\cite{novikov,chernyak}.

Proceeding in a standard way, we now compute  the hadronic side of the sum
rule. We can write down for $A(q_1^2,q_2^2,0)$ the following dispersion
relation:
\be
A(q_1^2,q_2^2,0)={1\over{\pi^2}} \int ds ds' {{\rho(s,s')}\over{(s-q_1^2)
(s'-q_2^2)}} \; .
\label{intg}
\ee
It should be observed that we have not written down in
(\ref{intg}) subtraction terms because, as proven in
 \cite{smilga1}, only
a subtraction polynomial $P_3(q_1^2,q_2^2)$ could be present in (\ref{intg}),
but it would vanish  after the Borel transform.

We divide the integration region in three parts \cite{noi}.
The first region (I) is the
square given by $ m_b^2 \le s,s' \le s_0$ and it
contains only the  $B^*$ pole, whose contribution is
\be
A_I(q_1^2,q_2^2,0)={{-2 \gibs f_{B^*}^2 m_{B^*}^2}
\over{f_{\pi}(q_1^2-m_{B^*}^2) (q_2^2-m_B^2)}}
\label{4}
\ee
where  $f_{B^*}$ is defined by
\be
<0|V_{\mu}(0)|B^*(\epsilon,p)>= \epsilon_{\mu} f_{B^*} m_{B^*}
\ee

The second (II) integration region is defined as follows:
$m_b^2 \le s \le s_0$ and $s'>s_0$ or $m_b^2 \le s' \le s_0$ and $s>s_0$.
Here we obtain a contribution coupling the vector current $V_{\mu}$ to the pion
and the $B^*$. Introducing the form factor $V$ as
\be
< \pi(q)| V_{\mu} | B^* (q_1, \epsilon) > \; = \; V(q_2^2)
\epsilon_{\mu\nu\alpha\beta} \epsilon^{\nu}
q_1^{\alpha} q^{\beta}
\label{6}
\ee
where $q_2=q_1-q$, we get
\be
A_{II}(q_1^2,q_2^2,0)= \fbs \mbs \left( \frac{V(q_2^2)}{q_1^2-\mbs^2} +
\frac{V(q_1^2)}{q_2^2-\mbs^2} \right)
\label{7}
\ee
In the previous formula one does not have to include the $B^*$ pole
contribution to $V(q^2)$, being already taken into account in $A_I$.
We assume that, taken away $B^*$, a single higher resonance of mass $m'$
contribute to $V$
\be
V_{res} (q^2) = \frac{k}{q^2-m'^2}
\label{8}
\ee
where $k$ is an unknown constant.

The third region is defined by $s,s' > s_0$, and under the assumption of
duality it should coincide with the asymptotic limit $q_1^2=q_2^2 \to -\infty$
in
(\ref{3}). One gets:
\be
A_{III}(q_1^2,q_1^2,0)= \frac{f_{\pi}}{q_1^2-s_0}
\label{10}
\ee

The hadronic side of the sum rule is the sum of the contributions from the
three regions:
\be
A_I(q_1^2,q_1^2,0)={{-2 \gibs f_{B^*}^2 m_{B^*}^2}
\over{f_{\pi}(q_1^2-m_{B^*}^2) (q_2^2-m_B^2)}} +
\frac{k'}{(q_1^2-\mbs^2)(q_1^2-m'^2)} +\frac{f_{\pi}}{q_1^2-s_0}
\label{11}
\ee
We have put $q_1=q_2$ and $k'= \mbs\fbs k$.

Equating now the hadronic and the QCD  sides of the sum rule, respectively
given by eq. (\ref{11}) and (\ref{3}),
and taking the Borel transform with parameter $M^2$ we find:
\bea
\frac{2 \gibs\mbs^2\fbs^2}{f_{\pi} M^2}& +& k' + \exp{(-\delta/M^2)}
 (f_{\pi}-k')=
\nn \\
&=& \exp{(\Omega/M^2)} \left[f_{\pi}-
\frac{1}{M^2}\left( {{2\qq m_b}\over{3 f_{\pi}}}+{{8 f_{\pi}
m_1^2}\over{9}} \right) \right. \nn\\
&-&\left. \frac{1}{M^4}\left(
 {{5 m_b^2 f_{\pi} m_1^2}\over{9}}-{{m_0^2\qq m_b}
\over{6 f_{\pi}}}\right)  +\frac{1}{M^6}
{{m_0^2\qq m_b^3}\over{6 f_{\pi}}}\right] \nn \\
& =& \exp{\Omega/M^2} S(M^2)
\label{12}
\eea
In the previous formula we have put $m'^2 \simeq s_0$ and we have introduced
the
parameters $\delta=s_0-\mbs^2$ and $\Omega=\mbs^2-m_b^2$.

Differentiating (\ref{12}) respect to the variable $1/M^2$ and combining the
first and second derivatives in order to eliminate the unknown parameter $k'$,
we obtain the following sum rule:
\bea
\gibs= \frac{f_{\pi}}{2 \mbs^2\fbs^2} \frac{\exp{\Omega/M^2}}{\delta} & &
\left[ \Omega (\Omega+\delta) S(M^2) +\right. \nn \\
& & \left. + (2 \Omega +\delta) \partial_{1/M^2} S(M^2) +\partial_{1/M^2}^2
S(M^2) \right]
\label{13}
\eea

To eliminate the parameter $k'$ one could also combine the first derivative
with the original sum rule (\ref{12}): we have checked that the two procedures
give the same numerical results. We have used the second derivative to make an
easy comparison with the sum rule for $\gib$ \cite{noi}:
\bea
\gib &= & \frac{4f_{\pi}m_b^2 \mbs}{2 m_B^3 f_B\fbs (3 \mbs^2+ m_B^2)}
\frac{\exp{(\Omega'/M^2)}}{(\delta'-\delta' \Delta_{B^*B}
/M^2 -\Delta_{B^*B})} \times \nn \\
 &\times &
\left[ \Omega' (\Omega'+\delta') S'(M^2) +
(2 \Omega' +\delta') \partial_{1/M^2} S'(M^2) +\partial_{1/M^2}^2
S'(M^2) \right]
\label{14}
\eea
where $\Delta_{B^*B}=
\mbs^2- m_B^2$, $\delta'= s_0 - m_B^2=\delta+\Delta_{B^*B}$,
 $\Omega'=
m_B^2-m_b^2=\Omega-\Delta_{B^*B}$ and
\bea
S'(M^2) &=&  f_{\pi} - {{\qq}\over{3 m_b f_{\pi}}}  \nn\\
&+&{1\over{M^2}} \lq -{{2\qq m_b}\over{3 f_{\pi}}}+{{10  f_{\pi} m_1^2}
\over{9}} + {{m_0^2\qq}\over{3 m_b f_{\pi}}} \rq \nn\\
&+& {1\over{2 M^4}} \lq -{{10 m_b^2 f_{\pi} m_1^2}\over{9}}+{{m_0^2\qq m_b}
\over{6 f_{\pi}}} \rq +\nn\\
&+& {{m_0^2\qq m_b^3}\over{6 f_{\pi} M^6}}
\label{14bis}
\eea

Eq. (\ref{14})  differs slightly from the one given in \cite{noi}, since it
keeps
track of the mass difference $\Delta_{B^* B}$.
The sum rules (\ref{13}) and (\ref{14}) have to be analyzed in the duality
region,
i.e. the region in $M^2$ where there exists a
 hierarchy among the different
contributions of higher dimension operators (this fixes the lower bound for
$M^2$ ); moreover we impose that the contribution of the parasitic term does
not exceed that of the resonance term, which fixes the upper bound in $M^2$. In
this
way we obtain for the $B$  $M^2$ in the range $20-40 \; GeV^2$ (for
$s_0=33-36 \; GeV^2$ ), and for the $D$  $M^2=4-7 \; GeV^2$ (for $s_0=6-8 \;
GeV^2$ ).
Using $m_b=4.6 \; GeV$ and
$m_c=1.34 \; GeV$  one gets:
\bea
f_{B^*}^2 \; \gibs & = &  0.0094 \pm 0.0018 \; GeV^2  \nn\\
f_{D^*}^2\; g_{D^*D^* \pi} & = &  0.017 \pm 0.004 \; GeV^2  \; ; \label{15}
\eea
and for the $\gi$ coupling
\bea
f_B \; f_{B^*} \; \gib & = &  0.0074 \pm 0.0014 \; GeV^2  \nn \\
 f_D \; f_{D^*}\; \gid & = &  0.0112 \pm 0.0030 \; GeV^2  \; ; \label{16}
\eea

Once multiplied by $2 m_P/f_{\pi}$ the figures in (\ref{16}) agree with those
given in \cite{noi}.

A recent calculation of the quantity reported in Eq. (\ref{16}) has been
given in Ref. \cite{Belyaev}. When expressed in our units
their results are as follows:
$f_B \; f_{B^*} \; \gib  =   0.0079 \pm 0.0007 \; GeV^2$ and
 $f_D \; f_{D^*}\; \gid  =   0.018 \pm 0.002 \; GeV^2$. The result for
the $B$  is only slightly larger than our outcome Eq. (\ref{16}), whereas
the result for the $D$ is significantly ($\simeq 60\%$) larger. The
origin of the discrepancy is in the different range of values for the Borel
parameter $M^2$, that, in the case of Ref. \cite{Belyaev}, are generally
smaller.  A possible origin of this difference is the fact that,
while
in this paper we use QCD sum rules in the soft pion limit,
in \cite{Belyaev} light cone
sum rules are adopted, which results in an expansion in operators of
increasing twist instead of increasing dimension. In particular we
have included a dimension 5 contribution which is proportional to the
$m_0^2\qq$
condensate. This term has no counterpart in \cite{Belyaev};
since it has to be kept small, its
inclusion in \cite{Belyaev}
 might result in a more stringent
 constraint on the  hierarchy among the
different contributions of the Operator Product Expansion, and, therefore,
in a more stringent lower limit on $M^2$.

We now expand the sum rules (\ref{13}) and (\ref{14}) in the parameter $1/m_Q$,
keeping the leading term and the first order corrections.
The leading term is the one surviving in the limit $m_b \to \infty$, and has
already been calculated in \cite {noi} for $\gi$.

To extract the $1/m_Q$ corrections we introduce
the following parameters, finite in the large mass limit:
\be
E=\frac{M^2}{2 m_b} \; ; \; y_0=\frac{s_0- m_b^2}{2 m_b} \; ;  \;
\omega=m_B-m_b
\label{17}
\ee
and the $1/m_Q$ corrections to the leptonic decay constants, $\gi$ and $\gis$
\be
f_M= \frac{\hat F}{\sqrt{m_Q}} \left( 1+ \frac{A}{m_Q} \right) \; \;~~~
f_{M^*}= \frac{\hat F}{\sqrt{m_Q}} \left( 1+ \frac{A'}{m_Q} \right)
\label{17a}
\ee
The coefficients $A$ and $A'$ have been computed in \cite{Ball} \cite{Neubert},
but only for the $B$ mesons. The corrections found are large and suffering of
large uncertainties: there are significant numerical differences
between
\cite{Ball} and \cite{Neubert}.

In the limit $m_Q \to \infty$ the right hand sides of (\ref{13}) and (\ref{14})
coincide (notice that $S=S'$ in this limit), confirming the result anticipated
in (\ref{equal}) and giving:
\be
g {\hat F}^2  =  \frac{f_{\pi} \exp{(\omega/E)}}{ (y_0 - \omega)} \left[
y_0 S_0(E) + (y_0 +\omega) \partial_{1/E}S_0(E) )+
\partial_{1/E}^2 S_0(E)\right]
\label{gasin}
\ee
where
\be
S_0(E) = f_{\pi} -{\qq\over{3f_{\pi} E}}-{{5f_{\pi} m_1^2}\over{36 E^2}} +
{{m_0^2\qq}\over{48 E^3 f_{\pi}}}
\ee
The duality region extends for $E=4-6 \; GeV$ and $y_0=1.1-1.3 \; GeV$.
Numerically one obtains:

\be {\hat F}^2 \; g = 0.040 \pm 0.005 GeV^3 \;
\label{gf}
\ee

While this result agrees within the error with that given in \cite{noi}, the
central value reported here is  $15 \;  \%$ larger due to a slightly different
choices of the phenomenological parameters.

We can then write the $1/m_Q$ expansion for the function $S$ and $S'$:
\be
S(E)= S_0(E) +\frac{S_1(E)}{m_Q} \; \; S'(E)= S_0(E) +\frac{S'_1(E)}{m_Q}
\label{18}
\ee
where
\bea
S_1(E) & = &  -{{4f_{\pi} m_1^2}\over{9 E}} +
{{m_0^2\qq}\over{24 E^2 f_{\pi}}}   \nn \\
S_1'(E) &= &   -{\qq\over{3f_{\pi} }}+{{5f_{\pi} m_1^2}\over{9 E}} +
{{m_0^2\qq}\over{48 E^2 f_{\pi}}}
\label{19}
\eea
{}From (\ref{13}) and (\ref{14}), one gets the following sum rules for the
parameters
$a$ and $b$:
\bea
a & = & \frac{f_{\pi} \exp{\omega/E}}{2 g {\hat F}^2 (y_0 - \omega)} \left[
(\omega^2+ 4\lambda)(y_0 S_0(E) +\partial_{1/E}S_0(E) )+ \right. \nn \\
&+& \left. 2 y_0 \omega S_1(E)+ 2(y_0+\omega) \partial_{1/E} S_1(E) +2
\partial_{1/E}^2 S_1(E) \right] + \nn \\
&+& \frac{\omega^2+4 \lambda}{2E} + \frac{\omega^2+4 \lambda}
{2(y_0 -\omega)} -2
(A'+\omega)
\label{20}
\eea
\bea
b & = & \frac{f_{\pi} \exp{\omega/E}}{2 g {\hat F}^2 (y_0 - \omega)} \left[
\omega^2 (y_0 S_0(E) +\partial_{1/E}S_0(E) )+ \right. \nn \\
&+& \left. 2 y_0 \omega S'_1(E)+ 2(y_0+\omega) \partial_{1/E} S'_1(E) +2
\partial_{1/E}^2 S'_1(E) \right] + \nn \\
&+& \frac{\omega^2+4 \lambda}{2E} + \frac{\omega^2+4 \lambda}{2(y_0 -\omega)} -
(A + A' +4 \omega)
\label{21}
\eea
where $\lambda$ has been given in (\ref{tbis}).
{}From the previous sum rules one gets:
\be
a+2A'= -0.15 \pm 0.20 \; GeV \;\;\;~~ \; b+A'+A= -1.15 \pm 0.20 \; GeV
\label{22}                                      \ee
and
\be
a-b+(A'-A)=0.99 \pm 0.02 \; GeV
\label{splitc}
\ee
Notice that the difference has a quite smaller uncertainty, due to a partial
cancellation of terms depending on the threshold.

Neglecting radiative corrections, $A$ and $A'$ are given by
\cite{Ball,Neubert}:
\be
A= -\omega + \frac{G_K}{2} + 3 G_{\Sigma} \;\;\; \;~~
A'= -\frac{\omega}{3} + \frac{G_K}{2} - G_{\Sigma}
\ee

Notice that the splitting of the couplings depends on the quantity $a-b$ that
contains only the difference $A'-A$ given by:
\be
A'-A = \frac{2}{3} \omega -4 G_{\Sigma}
\label{diff}
\ee

There is disagreement in the literature on the values of
the parameter $G_{\Sigma}$: at the $b$ quark
mass scale from ref.\cite{Ball} one gets $G_{\Sigma} = (0.042 \pm 0.034 \pm
0.023 \pm 0.030) \; GeV$, while in ref. \cite {Neubert}  the central value
$G_{\Sigma} \simeq -(0.052)\; GeV$ is quoted.
In view of this discrepancy, to provide an estimate of the difference
(\ref{diff}), we will approximate $A' -A \approx 2/3 \omega \approx 0.4 \;
GeV$,
obtaining
\be
a-b \approx 0.6 \;\; GeV
\label{ab}
\ee

\resection{Discussion and conclusions}

{}From (\ref{u}), (\ref{ab}) and
from the formula (\ref{split}) of the hyperfine mass splitting we
obtain:
\be
\Delta_B \approx g^2 (27.3 + 61.4 -75.8)~ MeV= 12.9 g^2 ~MeV
\label{z}
\ee
Notice that we have used in eq. (\ref{split}) $f=f_{\pi}=132 \; MeV$
for all the light pseudoscalar mesons of the octet. This is
suggested by the sum rule for $g$ which shows that $g/f$ is flavour
independent.
In eq. (\ref{z})
we have detailed the contributions $\Delta^0$, $\Delta^1$ and the one from
$\Delta_g/g$ respectively. We have also taken $\Lambda_{CSB}=1~GeV$.
 It is evident that there is a large cancellation among the last term and
the other ones. In order to be more quantitative we have to fix the value
of $g$, which, on the basis of our result (\ref{gf}), depends on the value
of ${\hat F}$. In Ref. \cite{noi} the range of values
$g \simeq 0.2-0.4$ was found; therefore, putting $g^2=0.1 $,
we would obtain
\be
\Delta_B \simeq 1.3 ~MeV
\label{z1}
\ee

The application of our results to the
charm case is more doubtful, in view of the large values of the $1/m_c$
correction $(a-b)/m_c$. By scaling the result (\ref{z1}) to the charm case,
one obtains
\be
\Delta_D = \frac{m_b}{m_c}\Delta_B \simeq 4.4 ~MeV
\label{z2}
\ee

In conclusion, our estimate of $\gis -\gi$ allows to include a previously
neglected term in the loop induced contribution to the hyperfine splitting.
Although our estimate is affected by an uncertainty in the value of
$G_{\Sigma}$, nevertheless this new term tends to cause
a substantial cancellation
and to reconcile the chiral calculation with the experimental data.
\par
\vspace*{1cm}
\noindent
{\bf Acknowledgements }\\

We would like to thank P. Colangelo for useful discussions.

\end{document}